\documentclass[10pt,twocolumn,superscriptaddress,nofootinbib]{revtex4-1}

\usepackage[scaled=1.03]{inconsolata} 
\usepackage[colorlinks=true, citecolor=blue, urlcolor=blue]{hyperref}  
\usepackage{graphicx} 
\usepackage[babel]{microtype}  
\usepackage{amsmath,amssymb,amsthm,bm,amsfonts,mathrsfs,bbm} 
\usepackage{xspace}  
\usepackage{pgfplots}
\newcommand{\tr}{\mbox{Tr}}

\newcommand{\oprd}[1]{|#1\rangle\langle#1|}

\newcommand{\ex}[1]{e^{#1}}

\begin{document}


\title{Thermodynamics of local baths in the context of work extraction}
\vspace{-5cm}
\hspace{5cm} YITP-17-95
\author{Tamal Guha}
\affiliation{Physics and Applied Mathematics Unit, Indian Statistical Institute,\\ 203, B. T. Road, Kolkata 700108 , India.}
\author{Mir Alimuddin}
\affiliation{Physics and Applied Mathematics Unit, Indian Statistical Institute,\\ 203, B. T. Road, Kolkata 700108 , India.}
\author{Amit Mukherjee}
\affiliation{Physics and Applied Mathematics Unit, Indian Statistical Institute,\\ 203, B. T. Road, Kolkata 700108 , India.}

\author{\\ Some Sankar Bhattacharya}
\affiliation{Physics and Applied Mathematics Unit, Indian Statistical Institute,\\ 203, B. T. Road, Kolkata 700108 , India.}

\author{Arup Roy}
\affiliation{Physics and Applied Mathematics Unit, Indian Statistical Institute,\\ 203, B. T. Road, Kolkata 700108 , India.}

\author{ Arpan Bhattacharyya}
\affiliation{Department of Physics and Center for Field Theory and Particle Physics, Fudan University, 220 Handan Road, 200433 Shanghai, P. R. China.}
\affiliation{
Center for Gravitational Physics, Yukawa Institute for Theoretical Physics,\\
Kyoto University, Kyoto 606-8502, Japan.}


\begin{abstract} 
We consider locally thermal states (for two qubits) with certain amount of quantum entanglement present between them. Unlike previous protocols we show how work can be extracted by performing local unitary operations on this state by allowing those two qubits to interact with  thermal baths of different temperatures, thereby gradually removing the entanglement  between them till they reach a direct product state. Also we demonstrate that, further work  can be extracted from  this direct product state by performing global unitary operation, thereby establishing that work can be extracted from a system composed of locally thermal subsystems even after removing quantum correlations between them if the subsystems are thermalized at different temperatures. Also we show that even if we consider a initial state where there is no entanglement between the two qubits, we can also extract work locally using our method. 
\end{abstract} 
\maketitle

\section{Introduction} \label{sec1}
In recent times, there is a surge of application of the idea of information in various branches of physics from condensed matter to quantum gravity,  especially in the context of  thermodynamics. Classical thermodynamics is perhaps one of the robust branches of physics, governed by its three laws. These laws can be derived from the microscopic physics of the underlying system. But there are many subtleties \cite{ref1,ref2,ref3,ref4} while applying these laws to finite systems where one cannot take  so called \emph{thermodynamic limit}. (i.e density of states tends to infinity smoothly with the increasing system size.) As a result, laws of thermodynamics take quite different form for the quantum systems mainly owing to the non-Gibbsian evolution of the initial state. Lots of effort have been put in the recent past to establish the laws of thermodynamics in the quantum regime. In this context, resource-theoretic aspects of quantum thermodynamics have developed. Although the importance of quantum correlations in the context of quantum thermodynamic is yet to be understood in full generality, but substantial amount of progress has been made in this direction recently \cite{review}. \par
One of the important area of research in quantum thermodynamics, is to determine how to extract work optimally from a quantum system. This is rather an old question starting from the work of \cite{Lifshitz'1978, Callen'1985}, but has been made more  concrete in recent past. In \cite{Pusz' 1978}, the mathematical framework of $C^*$-algebra has been used to study  the  question of optimal work extraction procedure from an isolated quantum system under \emph{cyclic Hamiltonian process}.  This question then later explored in Hilbert space formalism in \cite{Lenard'1978}. In this context, generally one start with  an isolated system and couple it with an external source for  short interval of time. This  is achieved by turning on a time dependent potential $V(t)$ which is non zero only between time $t=0$ and $t=\tau$ and then one study  the evolution of the system under the full hamiltonian 
\begin{equation}H(t)=H+ V(t),\end{equation} where $H$ is the intrinsic hamiltonian of the system under consideration. Without loss of any generality, it can take the following form 
\begin{equation} H=\sum_{i=1}^{N}\epsilon_{i}|i><i|, \quad \epsilon_{i+1}>\epsilon_i ,\end{equation} where the energy levels are nondegenerate and $N$ denotes the dimension of the underlying Hilbert space. Then the total amount of  extractable work ($W$) is simply the difference in internal energy between the final and initial state. 
\begin{equation}W=\tr[\rho(t=\tau)H]-\tr[\rho(t=0)H],
\end{equation} where, $\rho(t=\tau)= U^{\dagger} \rho(t=0) U$ and $ U=\ex{i\,\int_0^{\tau} H(t) dt }.$ So $\rho(t)$ and $\rho(t=0)$ shares the same spectrum and $U$ defines a global unitary operator. It has been further established in \cite{Pusz' 1978, Lenard'1978, Nieuwenhuizen'2004}, that  the maximum amount of work can be  extracted under this cyclic Hamiltonian process whenever the system evolves into a state, called \emph{passive} state ($\sigma$), from which no further work can be extracted. By the term \emph{passive} we mean, the underlying density matrix commutes with the Hamiltonian and the eigenvalues are non-increasing with the energy i.e, 
\begin{equation}\sigma= \sum_{i=1}^{N}s_{i} |i><i|,\quad s_{i+1}< s_{i}.
\end{equation}This maximal amount of work 
\begin{equation}\label{wmax} 
W_{max}=\tr[\rho(t=\tau)H]-\tr[\sigma H]
\end{equation}
(where $\sigma$ and $\rho$ have same eigenvalues), sometimes called as \emph{ergotropy} \cite{Nieuwenhuizen'2004} in the literature. Further, like in classical thermodynamics, for the individual quantum system there exists a protocol by which the amount of  work that can be extracted can be shown to be equal to the change in its free energy \cite{Popescu'2014}.  Apart from this, various works have been done to study the status of various laws of thermodynamics  for individual quantum systems \cite{Popescu'2014,Spekkens'2013, Horodecki'2013,Wehner'2014, Gour'2015}.\par 

But the question still remains,  is that all we can get?  Can one utilize  quantum correlations present in the system to further extract work?  In recent past this  topic of extracting work  utilizing correlations in multi-party quantum system has gained renewed interest \cite{Spekkens'2013, Horodecki'2013, Aberg'2013, Aberg'2014, Lostaglio'2015, Acin'2014}. In this context it seems underlying entanglement structure of the quantum system plays very crucial role. In \cite{Fannes'2013} it has been shown that, one can further extract work taking multiple copies of passive states and acting certain entangling unitary (these are global unitary operators ) on the state made up from these multiple copies $\sigma^{\otimes n }$, thereby beating the bound (\ref{wmax}) established in \cite{Nieuwenhuizen'2004}. This process is sometimes known as \emph{activation}.  But then this activation process doesnot work for the thermal states as they are completely passive state \cite{Pusz' 1978, passive}. Further, in \cite{Acin'2014}, the authors designed a protocol for a system composed of finitely many subsystems such that locally each of them are in thermal states  but the global state is an entangled state so that one can extract work \cite{Acin'2014}. To illustrate this, imagine a situation, where a bipartite state is shared between Alice and Bob initially, given by,
	\\
	\begin{equation} \label{ex}
	|\psi_{initial}\rangle= \frac{|00\rangle+ \ex{-\beta E }|11\rangle}{\sqrt{Z}}	
	\end{equation}
	with, Z= $1+\ex{-2\beta E}$.
	\\The local marginals of this state follows Normal probability distribution 
	($\frac{1}{Z},\frac{\exp{(-\beta E)}}{Z}$), which says that the local states are  {\it thermal}. So, it is impossible to extract any amount of work locally from this state. But one can use a global unitary on the joint state which can be lowered it to the lowest energy state \cite{Acin'2014}. Also the maximum amount of work that can be extracted using this global unitary is bounded by,
\begin{equation} \label{wmax1}
W_{max}\leq n E_{\beta},
\end{equation}
$\beta$ is the inverse temperature of the thermal bath with respect to which each of these subsystems are in thermal equillibrium and $n$ is the number of subsystem (for the above example $n$ is 2 and $E_{\beta}=E.$)
	\\ In this article we asked the question that, if Alice and Bob brought these two particles a distance apart, then can it be possible to extract some amount of work locally, but using  thermal baths of different temperatures? And we answer affirmatively. The dissipation of the correlation initially present between Alice and Bob due to the interaction with different temperature local baths is also studied.
	\\ Here we have considered a different kind of work extraction scenario, which is termed as {\it spontaneously extractable work}. The local thermal states of individuals will evolve by the interaction with local baths of different temperatures.  This evolution of the local states will decrease the free energy of the local particles, which can be stored in a work storage device ({\it storage battery}), as an amount of extracted work. So, one can switch off the bath Hamiltonian at any instant of time and the storage battery can be decoupled from the system with stored energy, which can be used further.
	\\ At $t\to \infty $ as the individual states acquire the local bath (different temperatures for both Alice and Bob) probability structure, so  they can't be correlated any more. Now relaxing the above mentioned spatially separated condition one can use a global bath to extract an amount of work from them jointly. 
The rest of the paper is organized as follows, in the section 2, we describe our protocol of extracting work by performing local unitary operations. In the section 3 we demonstrate how to extract work from a state which is a direct product of local thermal states with different temperatures by global unitary operations. Then in section 4 we summarize and conclude with some open questions. 

\section{Local Unitary operation and work extraction protocol: A dynamical setup} 
In this section we will discuss the in detail our method of extracting work by local unitary operations. We will show that even if we start with locally thermal state we can extract work from the joint state by local unitary operations. In the process we also demonstrate that, this method of spontaneously extracting work by local unitary has a virtue that, even if the joint state is a direct product state  of two local thermal we can still extract certain amount of work.  Below we explain this by considering three different examples. 

\subsection*{Case- I}  We  start with a two qubit entangled state described by the following density matrix,
	\begin{equation} \label{init}
\rho(t=0)=\frac{1}{Z} \left(
\begin{array}{cccc}
 1 & 0 & 0 & e^{-\beta  E} \\
 0 & 0 & 0 & 0 \\
 0 & 0 & 0 & 0 \\
 e^{-\beta  E} & 0 & 0 & e^{-2 \beta  E} \\
\end{array}
\right).
\end{equation}
Here $\beta =\frac{1}{T},$ $T$ is initial temperature characterizing this state. $Z=1+e^{-2\beta E}$ is the normalization such that, $\tr \rho=1.$ This two qubit system admits the following Hamiltonian,
\begin{equation} H=h_{w}\Big(Z_{a} \otimes I_{b}+I_{a}\otimes Z_{b}\Big),
\end{equation}
where, $a,b$ denote respectively the two particles under consideration. $h_{w}$ denotes the coupling strength. If there is no other interaction present, the evolution of the state,
\begin{equation}
\frac{d\rho(t)}{dt}=i[H,\rho(t)],
\end{equation}
cannot change its entanglement structure.  To change its entanglement structure we let the two particles to interact with the two thermal baths with respective temperature $\beta_1$ and $\beta_2. $ ($\beta > \beta_1, \beta_2$) We now add the appropriate interaction terms in the evolution equation which takes the following form \cite{book}, 
	\begin{align}
	\begin{split} \label{init1}
	\frac{d\rho(t)}{dt}=&i\,[H,\rho(t)]+\gamma_{1}(n_1+1) B+\gamma_{1} n_1 P\\&+\gamma_{2}(n_2+1) Q+\gamma_{2} n_2 R,
	\end{split}
	\end{align}
	The  interaction terms are defined in the following way \cite{book}, 
\begin{align}
	\begin{split}
&B=B_{1}.\rho(t).B_2-\frac{1}{2}B_2.B_1.\rho(t)-\frac{1}{2}\rho(t).B_2.B_1,\\&
P=B_2.\rho(t).B_1-\frac{1}{2}B_1.B_2.\rho(t)-\frac{1}{2}\rho(t).B_1.B_2,\\&
Q=C_1.\rho(t).C_2-\frac{1}{2}C_2.C_1.\rho(t)-\frac{1}{2}\rho(t).C_2.C_1\\&
R=C_2.\rho(t).C_1-\frac{1}{2}C_1.C_2.\rho(t)-\frac{1}{2}\rho(t).C_1.C_2.
\end{split}
	\end{align}
	where, \begin{align}
	\begin{split}
	&B_1=|0>_{a}{}_{a}<1|\otimes (|0>_{b}{}_{b}<0|+|1>_{b}{}_{b}<1|),\\&
	B_2=|1>_{a}{}_{a}<0|\otimes (|0>_{b}{}_{b}<0|+|1>_{b}{}_{b}<1|),\\&
	C_1=(|0>_{a}{}_{a}<0|+|1>_{a}{}_{a}<1|)\otimes |0>_{b}{}_{b}<1|,\\&
	C_2=(|0>_{a}{}_{a}<0|+|1>_{a}{}_{a}<1|)\otimes |1>_{b}{}_{b}<0|.
	\end{split}
	\end{align}
 $B,P,Q,R$ matrices are traceless. Now, because of this interaction, the initial entanglement will decrease with time and eventually it will go to zero as the two particles will thermalize with the their respective baths. This models resembles well known two-level system that we encounter often when we study dynamics of a system confined to a two dimensional subspace and the transitions to other level are negligible, e.g- laser systems. The equation (\ref{init1}) is nothing but a optical master equation of Lindblad type that is well studied in the literature \cite{book}. For all practical purpose we can choose \cite{book}, $\gamma_1=\gamma_2=\frac{4 \omega_0^3 |\vec d |^2}{3\bar h c^3}$ as the spontaneous emission rate where $\omega_0$ denotes the frequency of the transition between the two levels and $\vec d$  dipole operator. $n_1=n_1(\omega_0,\beta_1)$ and $n_2=n_2(\omega_0,\beta_2)$ denote the Planck distributions at this transition frequency with the corresponding temperatures $\beta_1$ and $\beta_2$ of the two baths. More details of this can be found in \cite{book}.
 
\begin{figure}
        \centering
        \includegraphics[width=.50\textwidth]{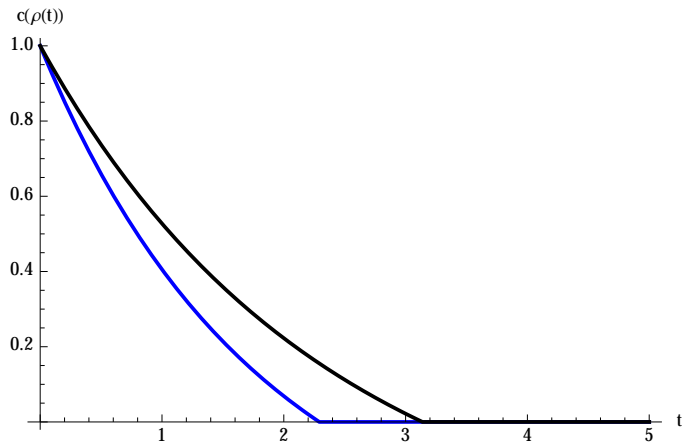}
        \caption{ Plot shows the decay of $C(\rho(t))$ with time $t.$The black plot is for $\beta E=.001, \beta_1  \omega_0=0.7, \beta_2 \omega_0=0.8$ and $\gamma_1=\gamma_2=0.105.$ The blue plot is for $\beta E=.001, \beta_1  \omega_0=0.5, \beta_2 \omega_0=0.6$ and $\gamma_1=\gamma_2=0.105.$
         All the quantities are in the natural unit i.e $\bar h=c=1.$ Also we have set $h_w=.1$  }
        \label{fig1}
\end{figure}

Next we solve (\ref{init1}) with the boundary condition (\ref{init}). Now $\rho(t)$ have to be hermitian and reduce to  $\rho(t=0)$ at $t=0.$ Imposing these two conditions and also using the fact that, the trace of $\rho(t)$ should remain  unity at every time we solve the above set of differential equations.  Below we quote the solution for $\rho(t).$ There are only two non-vanishing off-diagonal elements of $\rho(t)$. The density matrix takes the following form,
	\begin{equation} \label{init2}
\rho(t)=\frac{1}{Z'} \left(
\begin{array}{cccc}
 \rho_{1 1}(t)& 0 & 0 & \rho_{14}(t)\\
 0 & \rho_{22}(t) & 0 & 0 \\
 0 & 0 & \rho_{33}(t) & 0 \\
\rho_{41}^{*}(t) & 0 & 0 &\rho_{44}(t) \\
\end{array}
\right).
\end{equation}
 Also, $\frac{1}{Z'} (\rho_{11}(t)+\rho_{22}(t)+\rho_{33}(t)+\rho_{44}(t)=1.$ Detail expressions for these components  are given in the appendix. We first investigate how the entanglement changes with the time. With this density matrix (\ref{init2}) we compute the concurrence $ C(\rho(t))$  following \cite{wootters} as shown below.
\begin{eqnarray}
 C(\rho(t))=max\Big\{0,\lambda_1(t)-\lambda_2(t)-\lambda_3(t)-\lambda_4(t)\Big\},
\end{eqnarray}
 where, $\lambda_1(t), \lambda_2(t),\lambda_3(t),\lambda_4(t) $ are the eigenvalues of  
 \begin{equation}
 R(t) =\sqrt{\sqrt{\rho(t)}\tilde\rho(t)\sqrt{\rho(t)}},
 \end{equation} with,
 \begin{equation}
 \tilde \rho(t)= (\sigma_x \otimes \sigma_y)\rho(t)^{*}(\sigma_y\otimes \sigma_y), 
 \end{equation}
 in the decreasing order for all the values of $t$ under consideration.
$C(\rho(t)$ is monogamous and it gives us the measure how the entanglement of the initial state changes (decreases) with the time. We plot $C(\rho(t))$ for suitable values of the parameter as shown in the figure  (\ref{fig1}) and it shows the expected behaviour. As time increases it rapidly tends to zero, showing us that the  entanglement between these two particles goes to zero and the initial state reaches a direct product state.

   Given this scenario, we will focus how much work we can extract over time as the correlation being destroyed.  We  first write down the expression for the extractable work,
  \begin{eqnarray}
  W_{tot}=W_{a}+ W_{b},
  \end{eqnarray}
  $a$ and $ b$ denote the two particles.
  \begin{align}
  \begin{split}
 W_a=&\tr(\rho_{a}(t). H_a)-\tr( \rho_a(t=0). H_a)\\&-\frac{1}{\beta_1} \tr (\rho_{a}(t).\log (\rho_a(t)) -\rho_a(t=0). \log (\rho_{a}(t=0)),
   \end{split}
   \end{align}
 where, $\rho_a(t)=\tr_{b} \rho(t).$
  Similarly,
  \begin{align}
  \begin{split}
 W_a=&\tr(\rho_{b}(t). H_b)-\tr( \rho_b(t=0). H_b)\\&-\frac{1}{\beta_2} \tr (\rho_{b}(t).\log (\rho_b(t)) -\rho_b(t=0). \log (\rho_{b}(t=0)),
   \end{split}
   \end{align}
  where, $\rho_b(t)=\tr_a \rho(t)$ and 
\begin{align}
\begin{split}  
H_a=H_b=\left(
\begin{array}{cc}
 0 & 0 \\
 0 & 2 E \\\end{array}
\right)
\end{split}
\end{align}
  We then plot $W_{tot}$ with respect to $t.$ 
   \begin{figure}
        \centering
        \includegraphics[width=.50\textwidth]{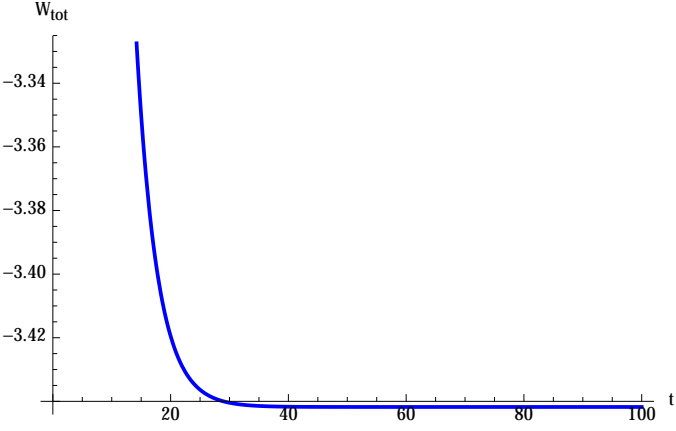}
        \caption{ Plot shows the decay of extractable work $W_{tot}$ with time $t.$ We have chosen $\beta=.1,\beta_1=.05,\beta_2=.03.$ All other quantities are expressed in the natural unit i.e $\bar h=c=1.$ $\gamma_1=\gamma_2=.105, h_w=.1,\beta_1 \omega_0=0.7, \beta_2 \omega_0=0.8$}
   \label{fig2}
\end{figure}
   From the figure~ (\ref{fig2}) it is evident over the time the amount of the extractable work decays which is what we expect as the correlation between the two particles also decays over time. Let us summarize what we have established here. We start with an entangled state of two particles system. We have allowed the two particles to interact with the two thermal baths of different  temperature thereby evolving the state to a state such that over the time initial entanglement between them decreases.  During this process we show that we can extract work till they get thermalized. We have performed  local unitary operation on a two particle entangled states to make  them a direct product state and extracted work in the process. At the end of the process the two particle gets thermalized with their respective bath. One last point to be noted  here that the amount of work that can be extracted not only depends on the initial $\beta$ and the bath temperatures (i.e $\beta_1, \beta_2 $) but also on the couplings $\gamma_1$ and $\gamma_2$. Here we have only studied the process for  some specific values of these parameters only just to demonstrate the basic features but we will leave the more systematic study of work extraction for all possible values of these parameters in near future. 
  \subsection*{Case-II}
  We consider yet another example.
  \begin{equation} \label{inita}
\rho(t=0)=\frac{1}{Z} \left(
\begin{array}{cccc}
 1 & 0 & 0 & 0 \\
 0 & 0 & 0 & 0 \\
 0 & 0 & 0 & 0 \\
 0& 0 & 0 & e^{-2 \beta  E} \\
\end{array}
\right).
\end{equation}
  From (\ref{init1}) we get the following, 
  \begin{equation} \label{init2a}
\rho(t)=\frac{1}{Z} \left(
\begin{array}{cccc}
  \rho_{1 1}(t)& 0 & 0 &0\\
 0 &  \rho_{22}(t) & 0 & 0 \\
 0 & 0 &\rho_{33}(t) & 0 \\
0& 0 & 0 & \rho_{44}(t) \\
\end{array}
\right),
\end{equation}
where the non zero components are same as that of (\ref{init2}) and the difference is that now we have only the  diagonal terms.  First of all note that from (\ref{init2a}), it is evident that, there is no correlation i.e $C(\rho(t))$ is zero. But still our methods enable us to extract work and we get the same features as shown in the figure (\ref{fig2}) for the previous case. We let this two qubits to interact with the two local bath with different temperatures ($\beta_1,\beta_2 \neq \beta$)  and in the process we can extract work till they get thermalized.
   \subsection*{Case-III}
   In this last example, we will start with a state which is a direct product of two thermal state (for simplicity we assume here that they are in same temperature but it can be generalized to other cases as well.).  So, 
  \begin{equation} \label{initb}
\rho(t=0)=\frac{1}{\hat Z} \left(
\begin{array}{cccc}
 1 & 0 & 0 & 0 \\
 0 & e^{- \beta  E} & 0 & 0 \\
 0 & 0 & e^{- \beta  E} & 0 \\
 0& 0 & 0 & e^{-2 \beta  E} \\
\end{array}
\right),
\end{equation}
where $\hat Z = (1+\exp^{-\beta E})^2.$  From (\ref{init1}) we get the following,
 \begin{equation} \label{init2c}
\rho(t)=\frac{1}{\hat Z} \left(
\begin{array}{cccc}
\hat  \rho_{1 1}(t)& 0 & 0 &0\\
 0 & \hat \rho_{22}(t) & 0 & 0 \\
 0 & 0 &\hat \rho_{33}(t) & 0 \\
0& 0 & 0 &\hat \rho_{44}(t) \\
\end{array}
\right),
\end{equation}
This again takes the same form as (\ref{init2a}) (again various components of this $\rho(t)$ are shown in the appendix). Then using the same procedure as before we can extract work and get a similar plot as in figure (\ref{fig2}). (notice that in this case also there is no correlation between the two local states.). \\
  We have demonstrated that if we start with locally thermal system, irrespective of the fact that, whether there exists correlation between them or not, we can extract work by letting the subsystems to interact with the local thermal baths of different temperatures until they thermalize. In the next section we demonstrate that further work can be extracted  from that final state but we have to use ``global" unitary instead of ``local" unitary. 
\section{Work extraction from states from two different thermal baths }
In this section we show that, even after the two particle thermalized with their respective baths with the corresponding (inverse) temperatures $\beta_1$ and $\beta_2$ ($\beta_1 \neq \beta_2$ is the crucial point here) it is possible to extract work, but unlike in the previous section where the work is extracted by local unitary operation, in this case we have to perform global unitary operation. Density matrix describing the particle which is in contact with the bath with inverse temperature $\beta_1$ , 
\begin{eqnarray} \label{t1}
\rho_{\beta_1}&=& p\oprd{0}+(1-p)\oprd{1}
\end{eqnarray}
where, $p=\frac{\ex{-\beta_1 E_0}}{\mathcal{Z}_1}=\frac{1}{\mathcal{Z}_1}$; $\mathcal{Z}_1=1+\ex{-\beta_1E}$, and $(1-p)=\frac{\ex{-\beta_1E}}{\mathcal{Z}_1}$. Here for simplicity  we have set $E_0=0.$ Similarly for the other particle in contact with the bath with inverse temperature $\beta_2 $ we have, 
\begin{eqnarray}
\rho_{\beta_2}&=& q\oprd{0}+(1-q)\oprd{1}
\end{eqnarray}
where, $q=\frac{\ex{-\beta_2 E_0}}{\mathcal{Z}_2}=\frac{1}{\mathcal{Z}_2}$; $\mathcal{Z}_2=1+\ex{-\beta_2E}$, and $(1-q)=\frac{\ex{-\beta_2E}}{\mathcal{Z}_2}$. Now that we have seen there is no correlation between them, then the resulting density matrix characterizing  the combined two particle system is,
\begin{equation} \rho_{\beta_1}\otimes \rho_{\beta_2}=\left( \begin{array}{cccc}
	pq & 0 & 0 & 0 \\
	0 & p(1-q) & 0 & 0 \\
	0 & 0 & q(1-p) & 0 \\
	0 & 0 & 0 & (1-p)(1-q)\end{array} \right). \label{t2}
	\end{equation}
	
	Following \cite{Fannes'2013}, there exists a  global unitary operation which can take this state to the following which has the same entropy as (\ref{t2}), 
	
\begin{equation} \label{t3}
\rho_{\tilde \beta}=k_1\oprd{00}+k_2\oprd{01}+k_3\oprd{10}+k_4\oprd{11}
\end{equation}
where,
\begin{eqnarray*}
k_1&=& \frac{1}{1+2\ex{-\tilde \beta E}+ \ex{-2\tilde \beta E}}\\
k_2=k_3&=& \frac{\ex{-\tilde \beta E}}{1+2\ex{-\tilde \beta E}+ \ex{-2\tilde \beta E}}\\
k_4&=& \frac{\ex{-2\tilde \beta E}}{1+2\ex{-\tilde \beta E}+ \ex{-2\tilde \beta E}}
\end{eqnarray*}
The entropy for (\ref{t3}) is given as, 
let, $x=\ex{-\tilde \beta E},~x^2=\ex{-2\tilde \beta E}$ 
\begin{eqnarray*}
S(\rho_{\tilde \beta}) &=& \frac{1}{1+2\ex{-\tilde \beta E}+ \ex{-2\tilde \beta E}} \log\left(\frac{1}{1+2\ex{-\tilde \beta E}+ \ex{-2\tilde \beta E}} \right)\\
&&+  \frac{2\ex{-\tilde \beta E}}{1+2\ex{-\tilde \beta E}+ \ex{-2\tilde \beta E}}\log\left(\frac{\ex{-\tilde \beta E}}{1+2\ex{-\tilde \beta E}+ \ex{-2\tilde \beta E}} \right)\\
&&+  \frac{\ex{-2\tilde \beta E}}{1+2\ex{-\tilde \beta E}+ \ex{-2\tilde \beta E}}\log\left(\frac{\ex{-2\tilde \beta E}}{1+2\ex{-\tilde \beta E}+ \ex{-2\tilde \beta E}} \right)\\
&=& 2\left( \log (1+x)-\frac{x}{1+x}\log x\right)\\
&=& 2\mathcal{H}(\frac{1}{1+x})=2\mathcal{H}(k).
\end{eqnarray*}
Also the entropy for the state mentioned in (\ref{t2}) is given as ,
\begin{eqnarray}
S(\rho_{\beta_1}\otimes \rho_{\beta_2})&=& \mathcal{H}\left[pq,p(1-q),q(1-p),(1-p)(1-q)\right]\nonumber\\
&=& p \log p+q\log q+(1-p)\log (1-p)\nonumber\\ & &+(1-q)\log (1-q)\nonumber\\
&=& \mathcal{H}(p)+\mathcal{H}(q)\label{aa}.
\end{eqnarray}	
As they are equal,
\begin{eqnarray}\label{t4}
	2\mathcal{H}(k)&=& [\mathcal{H}(p)+\mathcal{H}(q)] \nonumber \\ 
	2\mathcal{H}\left(\frac{1}{1+\ex{-\tilde \beta E}} \right) &=& \left[\mathcal{H}\left(\frac{1}{1+\ex{-\beta_1 E}} \right) +\mathcal{H}\left( \frac{1}{1+\ex{-\beta_2E}}\right) \right]\nonumber \\
	\end{eqnarray}
	So the amount of extractable work will simply be equal to the difference between the  free energy of these two states. 
	\begin{eqnarray} \label{t5}
	W=E(\rho_{\tilde \beta})-E(\rho_{\beta_1}\otimes \rho_{\beta_2})\nonumber\\
	= E \left[\frac{2\ex{-\tilde \beta E}}{1+\ex{-\tilde \beta E}}-\frac{\ex{-\beta_1 E}}{1+\ex{-\beta_1 E}}-\frac{\ex{-\beta_2 E}}{1+\ex{-\beta_2 E}} \right] 	
	\end{eqnarray}
	Now we  show that under which condition the  work defined in (\ref{t5}) is positive. We notice that, if $\beta_1>\beta_2$,
	\begin{eqnarray}
	\frac{1}{1+\ex{-\beta_1 E}}&>&\frac{1}{1+\ex{-\beta_2 E}}.
	\end{eqnarray}
	From this it follows, 
	\begin{eqnarray} \label{t6}
	\mathcal{H}\left( \frac{1}{1+\ex{-\beta_1 E}}\right) &<& \mathcal{H}\left( \frac{1}{1+\ex{-\beta_2 E}}\right) 
	\end{eqnarray}
	Combining (\ref{t4}) and (\ref{t6}) we get, 
	\begin{eqnarray}
	H\left( \frac{1}{1+\ex{-\tilde \beta E}}\right) &<& H\left( \frac{1}{1+\ex{-\beta_2 E}}\right) 
	\end{eqnarray}	
	This further implies,
	\begin{equation}\tilde \beta>\beta_2.
	\end{equation}
	And using this fact it can be easily shown that, $W$ defined in (\ref{t5}) is positive. Hence we prove that by performing global unitary operation on the two particle states where each of the individual particle are in a thermal states with different temperature one can extract work.  This establishes our claim that we made at the beginning: \emph {it is possible to extract work locally by allowing the subsystems to interact with thermal baths of different temperatures (even if there are no correlations between different subsystems)  till all the correlations between them are being removed  and after that, further work can be extracted by using global unitary.}

\section{Summary and Discussion}	
In this work, we have explored yet another avenue of the work storing capacity of the system in presence (or absence)  of quantum correlations. First, we have considered a 2 qubit system where each of the two qubits are in a local thermal state (characterized  by the  inverse  temperature $\beta,$ which can be arbitrary), but the global state is not a direct product sate. We have shown that given this scenario we can perform local unitary operations unlike the global unitary operations considered perviously in the literature to extract some amount of work. We  let these two qubits to interact with two thermal bath with different temperatures (inverse) $\beta_1$ and $\beta_2$ ( $< \beta$) and let them thermalize over time. In the process the correlation between them are being removed and we can extract work. One important point to be noted that, it has been shown previously  (for e.g in \cite{Acin'2014}) that for local thermal state once the correlation is completely removed one cannot further extract any work ``globally" as the resulting state becomes completely passive. We circumvent this situation in our protocol by allowing the two qubits to thermalize ``locally". Then we demonstrated that it is further possible to extract work by global unitary operations from this resulting state. Then we consider examples of states where there is no entanglement between the two qubits. But we have shown that, still we will be able to extract work from that state again by local unitary operations. So we basically established a protocol which is a combination of local and global unitary operations and gives a way to extract work locally form a system composed of local thermal sub systems irrespective of any quantum correlation present between them. It will be an interesting future study to compare the amount of the extractable work from this protocol with the bound established in \cite{Acin'2014} and investigate what is the maximum possible work that can be extracted using our protocol. Also it will be interesting to generalize this method for n-qubit systems which will give us intuition how to apply our protocol for more realistic open quantum systems. Also it is possible to study and apply this for non equilibrium scenario \cite{None} and investigate the role of fluctuation- dissipation theorem for our protocol \cite{fluc1,fluc2, fluc3, fluc4} and many more. We hope to get back to these problems in near future. \\

 \section{Acknowledgements}
 AB like to thank Prof. Ling-Yan Hung for her constant support and encouragement and organizing the conference  ``Quantum Entanglement 2017 " jointly with Prof. Feng-Li Lin hosted by Fudan University and National Taiwan University (NTU) where beautiful lectures on quantum thermodynamics were being delivered.  AB also like to thank Prof. Guruprasad Kar  for valuable discussions and hosting during his visit to the Physics and Applied Mathematics Unit (PAMU), Indian Statistical Institute (ISI), Kolkata where this work was initiated. AB acknowledges support from Fudan University and Thousand Young Talents Program. AB also acknowledges support form JSPS fellowship (P17023) and Yukawa Institute of Theoretical Physics (and mentor Prof. Tadashi Takayanagi) where the project has been concluded.  AM acknowledges support from the CSIR
project 09/093(0148)/2012-EMR-I. MA acknowledges support from the CSIR project 09/093(0170)/2016-EMR-I. All authors would like to thank Prof. Sibasish Ghosh for many  fruitful suggestions and sharing their upcoming work which is relevant for our study. \par
\vspace{0.5cm}
{\bf Note Added:} While preparing the draft we became aware of the result \cite{sg}, which also stands as a strong evidence for our work and further motivates us to extend our study for various other important scenario in future. 
\appendix
\section{}
  Here we provide the detailed expressions for the components of the density matrix $\rho(t)$ as mentioned in (\ref{init2}).
 \begin{align}
 \begin{split}
\rho_{14}(t)=\rho^{*}_{41}(t)&=\frac{e^{-\beta E } e^{-t \left(\gamma _1 \left(n_1+0.5\right)+\gamma _2 \left(n_2+0.5\right)\right)-4 i t h_w}}{Z},
 \end{split}
\end{align}
where $(*)$ denotes the complex conjugate. $\rho(t)$ has 4 diagonal elements also.
\begin{align}
\begin{split}
&\rho_{11}(t)=\frac{1}{(1+2 n_1)(1+2 n_2)Z}e^{-2  \beta E } \\&e^{-t \left(\gamma _1+\gamma _2+2 \gamma _1 n_1+2 \gamma _2 n_2\right)} \Big(e^{2 e \beta } \left(\left(n_1+1\right) e^{\gamma _1 \left(2 n_1+1\right) t}+n_1\right)\\& \left(\left(n_2+1\right) e^{\gamma _2 \left(2 n_2+1\right) t}+n_2\right)+\left(n_1+1\right) \left(n_2+1\right)\\& \left(e^{\gamma _1 \left(2 n_1+1\right) t}-1\right) \left(e^{\gamma _2 \left(2 n_2+1\right) t}-1\right)\Big),
\end{split}
\end{align} 
\begin{align}
\begin{split}
&\rho_{22}(t)=\frac{1}{\left(2 n_1+1\right) \left(2 n_2+1\right) Z}e^{-2  \beta  E} \\&\Big(\left(n_1+1\right) n_2 \left(e^{2  \beta E}+1\right)+n_2 \left(n_1 \left(e^{2  \beta E }-1\right)-1\right)\\& e^{-t \left(\gamma _1+2 \gamma _1 n_1\right)}-\left(n_1+1\right) \left(n_2 \left(e^{2  \beta E }-1\right)-1\right)\\& e^{-t \left(\gamma _2+2 \gamma _2 n_2\right)}-\left(n_1 \left(n_2 e^{2  \beta E }+n_2+1\right)+n_2+1\right) \\&e^{-t \left(\gamma _1+\gamma _2+2 \gamma _1 n_1+2 \gamma _2 n_2\right)}\Big),
\\&\rho_{33}(t)=\frac{1}{\left(2 n_1+1\right) \left(2 n_2+1\right) Z}e^{-2  \beta  E} \\&\Big(n_1 \left(n_2+1\right) \left(e^{2  \beta E }+1\right)-\left(n_2+1\right) \left(n_1 \left(e^{2  \beta E}-1\right)-1\right)\\& e^{-t \left(\gamma _1+2 \gamma _1 n_1\right)}+n_1 \left(n_2 \left(e^{2  \beta E}-1\right)-1\right) \\&e^{-t \left(\gamma _2+2 \gamma _2 n_2\right)}-\left(n_1 \left(n_2 e^{2  \beta E}+n_2+1\right)+n_2+1\right) \\&e^{-t \left(\gamma _1+\gamma _2+2 \gamma _1 n_1+2 \gamma _2 n_2\right)}\Big),
\end{split}
\end{align} 
\begin{align}
\begin{split}
&\rho_{44}(t)=\frac{1}{\left(2 n_1+1\right) \left(2 n_2+1\right) Z}e^{-2  \beta E } e^{-t \left(\gamma _1+\gamma _2+2 \gamma _1 n_1+2 \gamma _2 n_2\right)} \\& \Big(n_1 \Big(n_2 e^{2  \beta  E}-n_2 \left(e^{2  \beta  E}-1\right) e^{\gamma _2 \left(2 n_2+1\right) t}+n_2 \left(e^{2  \beta E }+1\right) \\&e^{t \left(\gamma _1+\gamma _2+2 \gamma _1 n_1+2 \gamma _2 n_2\right)}+\left(-n_2 e^{2  \beta E }+n_2+1\right) e^{\gamma _1 \left(2 n_1+1\right) t}\\&+n_2+1\Big)+n_2 e^{\gamma _2 \left(2 n_2+1\right) t}+n_2+1\Big).
\end{split}
\end{align} 
Next we give all the components of the density matrix as given in (\ref{init2c}) below. 
\begin{align}
 \begin{split}
\hat \rho_{11} (t)= &\frac{1}{(2 n_{1}+1) (2 n_{2}+1) \hat Z }e^{-2 \beta E-t (\gamma_{1}+\gamma _{2}+2 \gamma_{1} n_{1}+2 \gamma_{2} n_{2})}\\&
\Big(e^{\beta E} \left((n_{1}+1) e^{\gamma_{1} (2 n_{1}+1) t}+n_{1}\right)\\&+(n_{1}+1) \left(e^{\gamma_{1} (2 n_{1}+1) t}-1\right)\Big) \Big(e^{\beta E} \Big((n_{2}+1) e^{\gamma_{2} (2 n_{2}+1) t}\\&+n_{2}\Big)+(n_{2}+1) \left(e^{\gamma_{2} (2 n_{2}+1) t}-1\right)\Big), \end{split}
\end{align}
\begin{align}
\begin{split}
\hat \rho_{22}(t)=&\frac{1}{(2 n_{1}+1) (2 n_{2}+1) \hat Z}e^{-2 \beta E1-t (\gamma_{1}+\gamma_{2}+2 \gamma_{1} n_{1}+2 \gamma_{2} n_{2})}\\&
\Big(e^{\beta E} \left((n_{1}+1) e^{\gamma_{1} (2 n_{1}+1) t}+n_{1}\right)\\&+(n_{1}+1) \left(e^{\gamma_{1} (2 n_{1}+1) t}-1\right)\Big)\Big(-e^{\beta E} n_{2}+\left(e^{\beta E}+1\right)\\& n_{2} e^{\gamma _{2} (2 n_{2}+1) t}+n_{2}+1\Big), \end{split}
\end{align}
\begin{align}
\begin{split}
\hat \rho_{33}(t)=&\frac{1}{(2 n_{1}+1) (2 n_{2}+1) \hat Z }e^{-2\beta E-t (\gamma_{1}+\gamma_{2}+2 \gamma_{1} n_{1}+2 \gamma_{2} n_{2})}\\&
\Big(n_{1} \Big(-e^{\beta E}+e^{\beta E+2 \gamma_{1} n_{1} t+\gamma_{1} t}+e^{\gamma_{1} (2 n_{1}+1) t}+1\Big)+1\Big)\\& \Big(e^{\beta E} \left((n_{2}+1) e^{\gamma_{2} (2 n_{2}+1) t}+n_{2}\right)+(n_{2}+1)\\& \left(e^{\gamma_{2} (2 n_{2}+1) t}-1\right)\Big), \\
\hat \rho_{44}(t)=&\frac{1}{(2 n_{1}+1) (2 n_{2}+1) \hat Z }e^{-2\beta E-t (\gamma_{1}+\gamma_{2}+2 \gamma_{1} n_{1}+2 \gamma_{2} n_{2})}\\&\Big(n_{1} \left(-e^{\beta E}+e^{\beta E+2 \gamma _{1} n_{1} t+\gamma _{1} t}+e^{\gamma_{1} (2 n_{1}+1) t}+1\right)+1\Big) \\&\Big(n_{2} \left(-e^{\beta E}+e^{\beta E+2 \gamma _{2} n_{2} t+\gamma_{2} t}+e^{\gamma_{2} (2 n_{2}+1) t}+1\right)+1\Big).
\end{split}
\end{align}

\end{document}